\documentclass{raa}           
\usepackage{graphicx,times}
\usepackage{natbib}
\usepackage{amssymb,amsmath}
\bibpunct{(}{)}{;}{a}{}{,}

\usepackage[T1]{fontenc}
\usepackage{tikz}
\usetikzlibrary{shapes.geometric}
\usetikzlibrary{arrows,calc,positioning}

\tikzstyle{intg}=[draw,minimum size=3em,text centered,text width=6.cm]
\tikzstyle{decision} = [diamond, draw, thick,
text width=4.5em, text badly centered, inner sep=1pt]

\usepackage[a4paper=true,pagebackref=true,driverfallback=dvipdfm]{hyperref}
\hypersetup{pdftitle = The title of my PDF, pdfauthor = My name, pdfsubject= The subject, pdfkeywords = keyword1 keyword2 keyword3} 
\hypersetup{colorlinks = true, linkcolor = green, anchorcolor = red, citecolor = blue, filecolor = red, pagecolor = red, urlcolor = red}

\begin{document}

   \title{Measurements of atmospheric turbulence parameters at Vainu Bappu Observatory using short-exposure CCD images}

 \volnopage{ {\bf 20XX} Vol.\ {\bf X} No. {\bf XX}, 000--000}
   \setcounter{page}{1}
   
   \author{Sreekanth Reddy V\inst{1}, Ravinder Kumar Banyal\inst{1}, Sridharan R\inst{1}, Aishwarya Selvaraj\inst{1}
   }

   \institute{ Indian Institute of Astrophysics, Bangalore 560034, 
India; {\it sreekanth@iiap.res.in}\\
\vs \no
   {\small Received 20XX Month Day; accepted 20XX Month Day}}

\abstract{We report the atmospheric turbulence parameters namely, atmospheric seeing, the tilt-anisoplanatic angle ($\theta_0$) and the coherence time ($\tau_0$), measured under various sky conditions, at Vainu Bappu Observatory in Kavalur. Bursts of short exposure images of selected stars were recorded with a high-speed, frame-transfer CCD  mounted on the Cassegrain focus of a newly commissioned 1.3 m telescope. The estimated median seeing is $ \approx 1.85^{\prime\prime} $ at wavelength of $\sim 600$ nm, the image motion correlation between different pairs of stars is $\sim 44\%$  for $\theta_0 \approx 36^{\prime\prime}$ and mean $\tau_0$ is $\approx 2.4$ ms. This work was motivated by the design considerations and expected performance of an adaptive optics system that is currently being planned for the telescope. 
}

    \authorrunning{V. S. Reddy et al.}            
    \titlerunning{Measurement of atmospheric turbulence parameters} 
    \maketitle

%
\section{Introduction}
\label{sect:intro}
The wave-front of  light beam arriving from  distant astronomical sources is aberrated after passing through the turbulent layers of the Earth's atmosphere. The spatial and temporal inhomogeneities in the refractive-index of the air along the beam path produce random phase perturbations, impairing the performance of ground-based telescopes. The influence of wave-front distortion on optical system was investigated by \cite{fri65,Fried66} using Kolmogorov's model of turbulence. The impact of spatial structure of atmospheric turbulence is described by Fried's parameter $r_0$, a standard measure of atmospheric seeing.

Normally, the image resolution varies depending on the seeing $r_0$, telescope aperture size $D$ and exposure time $t$. For $D/r_0>1$ and central wavelength $\lambda_0$, the angular resolution of the telescope is limited to $\sim\lambda_0/r_0$ as opposed to diffraction limited case of $\sim\lambda_0/D$. Further, for seeing-limited observations the telescope sensitivity (number of photons received per unit area in the detector plane) scales with aperture as $D^2$, unlike  aberration-free case where sensitivity (\citealt{har98}) is proportional to $D^4$.

\begin{figure}
    \centering
	\includegraphics[height=8.5cm]{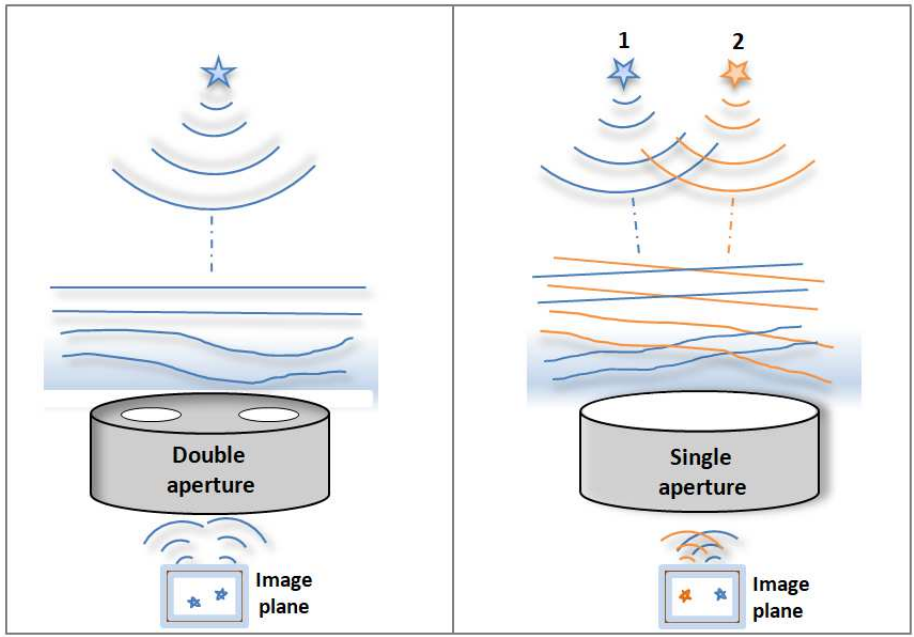}
    \caption{\label{fig:fig1} 
    Illustration of measuring differential tilt of a stellar wave-front sampled at different locations by two sub-aperture (left-hand panel) and differential tilt between a pair of stars within the field of a mono-pupil telescope (right-hand panel).}
\end{figure}

The wave-front aberrations across the entrance pupil of a telescope are distributed over large range of spatial frequencies. The lowest order aberration is the wave-front tilt which leads to overall angle-of-arrival fluctuations at the aperture plane.  This time-varying tilt  is responsible for short-exposure image motion (hereafter `\emph{image motion}' means the motion of the centroid of star image)  at the detector plane. \cite{fri75} had derived the expression for the mean-square difference in angle-of-arrival of light seen by two small sub-apertures separated by certain distance that sample the wave-front of a single star as shown in Figure~\ref{fig:fig1}.  This forms the basis for most contemporary seeing measurements carried out with differential image motion monitors (\citealt{sar90,wil99, tok02}).

Full potential of a ground telescope is realized by an adaptive optics system which measures the wave-front errors and applies corrections to compensate the tip-tilt as well as the high-order aberrations which cause the blurring.

The wave-front sensing is done either on a bright science target or with a guide star --a natural bright star or a laser  beacon created in the close vicinity of faint science target. The wave-front aberrations can differ significantly as the light from guide star and science object take slightly different paths along the atmosphere. The decorrelation in wave-front errors along two propagation paths is a measure of anisoplanatism. The high order aberrations usually correlate well only over a small angular field ($\lesssim 10^{\prime\prime}$),  while the correlation  for lowest order tip-tilt typically extends beyond several 10s of arc seconds. The degree of correlation is again determined by the atmospheric seeing and angular separation (\citealt{mcc91}).  The variance of wave-front tilt difference from two stars across a common telescope aperture, as illustrated in Figure~\ref{fig:fig1}, is generally referred to as \emph{tilt- anisoplanatism}.

Previous attempts to measure tilt-anisoplanatism used binary stars at various angular separations. Those observations suffered from restricted bandwidth limitations and lack of bright star pairs at specific angular separations. An early measurement of tilt-anisoplanatism was reported  by \cite{teo88} with 1.2 m Firepond facility at Lincoln Laboratory. In that experiment two quad trackers measured the differential tilt between three binary star pairs with separation varying from  $10^{\prime\prime}$ to $22^{\prime\prime}$. 

Subsequently, \cite{siv95} used fast readout CCD to capture multiple stars inside a single frame. A linear rise in standard deviation of differential tilt was observed for star separation varying from $10^{\prime\prime}- 58^{\prime\prime}$. In another experiment images of the moon's edge were used to obtain statistical estimate of tilt angular correlation and tilt averaging functions over a continuous wide range of angular separations (\citealt{bel97}).

The turbulence also produces temporal phase fluctuations along with spatial distortions in wave-front. These fluctuations are characterized by atmospheric coherence time ($\tau_{0}$). Within this time limit the phase fluctuations are negligible. In AO systems, time interval between wave-front sensing and correction should be constrained within unit $\tau_0$. Thus, it requires a temporal closed loop bandwidth (BW) $>>{1}/{2\pi \tau_0}$.   
 
As $\tau_0$ is dependent on $r_0$ and wind velocity, its measurement using short exposure images is influenced by data sampling time (\citealt{har98}). Because of this, different definitions of coherence time of phase fluctuations given by \cite{breck94}, differ by a factor as large as 7. Based on these approximations, \cite{dav96} have explained the effect of data sampling time on the measurement of turbulence. These earlier analysis have mentioned the optimum sampling time in the order of few milli seconds. 

These experiments emphasize the usefulness of fast, short exposure images of a star in determining crucial atmospheric turbulence parameters such as seeing ($\lambda/r_0$) and atmospheric time constant ($\tau_0$) and several pairs of stars can be used to measure tilt-anisoplanatic angle ($\theta_0$) (\citealt{mar87,dav96,kel07}).

We used short exposure images for the estimation of the atmospheric turbulence parameters, obtained with 1.3 m telescope at Kavalur Observatory. Short exposure images of selected stars were taken with a high-speed CCD camera. 

We exploited  the ability to define multiple region of interest (MROI) within the CCD frame to record simultaneous position measurements of stellar images with large angular separation up to $\approx 212^{\prime\prime}$, at the image plane for measurement of $\theta_0$.

The remaining material in the paper is organized as follows. The details of the high speed imaging camera, Target selection criteria, observation methodology and data analysis are described in Section 2.  In Section 3, we describe the seeing measured from the $rms$ image motion and compare it with that measured from other methods.  In Section 4, we describe the measurement of tilt-anisoplanatic angle and, in Section 5, we describe the estimation of the atmospheric coherence time $\tau_0$.   Finally, in section 6, we discuss relevance of these parameters in designing an AO system for the telescope. In Section 7 we summarize the results.

\section{Target selection and Observational  Methodology }

The 1.3 m telescope at Vainu Bappu Observatory, Kavalur, is newly commissioned in the year 2014. It is a Ritchey-Chretien telescope with hyperbolic primary and secondary mirrors. The telescope is located at an elevation of 750 m above the mean sea level at Javadi hills in southern India at $78^\circ 50^\prime E$ and $12^\circ 34^\prime N$. As the telescope is user friendly for operation, has better tracking and better optical quality, it is chosen for technology demonstration of AO system. Prior to the development of an AO system, on-site estimation of turbulence parameters could provide crucial information for the design of such system. Thus, a study of turbulence parameters is started on-site of 1.3 m telescope during the 1st quarter of 2016. The relevant specifications of the telescope are given in Table~\ref{tab:Telescope}.

\begin{table}
    \bc
	\centering
	\caption{Specifications of J. C .Bhattacharya telescope and CCD}
	\label{tab:Telescope}
	\begin{tabular}{lccr} 
		\hline\noalign{\smallskip}
		Property & Value \\
		\hline\noalign{\smallskip}
		Primary aperture diameter& 1.3 m \\
		Focal length & 10.4 m \\
		F ratio & 8 \\
		Central obscuration & 42 cm \\
		Mount type & equatorial \\
		Camera & Princeton Instruments ProEM xEcelon \\
		Detector size & 1024 x 1024 \\
		Pixel size & $13 \mu$m \\
		Pixel scale & $0.26^{\prime\prime}$ \\
		FOV & $ \approx 4^{\prime} \times 4^{\prime}$ \\
		Gain & 1.37 $e^-$/ADU \\
		Read Noise & 10.35 $e^-$ \\
		Bit depth & 16 bit/pixel \\
		Readout type & Frame transfer \\
		Readout rate & 10 MHz \\
		\noalign{\smallskip}\hline
	\end{tabular}
	\ec
\end{table}

\subsection{High speed CCD camera}
The rapid imaging requires a high speed CCD. It is desired to have continuous exposure with minimum time lag between two successive frames.  For tilt isoplanatic angle measurements two or more objects need to be observed simultaneously in a given field. The lower and upper limit of angular separation between two objects is set by the resolvability between the two objects and the CCD size respectively. The large separation between the objects occupies larger frame size and thus requires more readout time. To avoid reading unnecessary pixels, new technology also permits the selection of multiple ROIs on CCD. This useful functionality can further reduce the readout time. 

We used Princeton Instruments ProEm eXcelon ($1024\times1024$, pixel size = $13 \mu$m) EMCCD for our observations. The CCD used frame transfer mode for continuous exposure and simultaneous readout. In this mode, the detector has active and masked areas. After the exposure, the data is vertically (parallel) shifted from active area to masked area. This shift occures within few micro seconds. Thus, the active area is immediately available for next exposure (\citealt{ProEMManual}) while the image is read out from the masked section. The time lag between two successive frames is $0.8 \mu$sec. The current version of the CCD can readout the data with 10 MHz clock speed. It has quantum efficiency (QE)  $\approx90\%$ for visible band of wavelength (\citealt{PICCD}). The full frame of the CCD cover $\approx 4^{\prime} \times 4^{\prime}$ on-sky Field of View (FoV) with plate scale of $0.26^{\prime\prime}$/pixel. The CCD has an option to capture multiple ROIs within a single frame with user defined window size for each. The CCD specifications are given in Table~\ref{tab:Telescope}. 
\subsection{Observation methodology}
	
The telescope is used for intensive observations in rapid imaging mode. For the operation, target selection with respect to their magnitude, hour angle and declination is crucial. These variables influence the estimation of turbulence parameters.  

A list of target was chosen that satisfy the following criteria. The targets should be bright enough to have sufficient signal to noise ratio (SNR), pair should be spatially resolved and fall within the CCD frame. Thus, multiple pair of stars with  magnitude up to $m_v \approx 8$ and  on-sky angular separation  ranging  from  $6.4^{\prime\prime}-212^{\prime\prime}$ were chosen. The target objects are listed from Harvard revised bright star catalogue (\citealt{Hrobj64}). The list of observed objects (brighter companion in each pair) is compiled in Table~\ref{tab:Targets}.

\begin{table}
	\centering
	\caption{List of observed stars with varying angular separation. In this table, 'Dec' is declination of the object, 'RA' is right ascension,'$m_v$' is apparent magnitude, '$\Delta m_v$' is magnitude different between two objects and 'sep' is angular separation between the objects in arc seconds.}
	\label{tab:Targets}
	\begin{tabular}{lccccr} 
	      \hline
	      Object&Dec&RA&$m_v$& $\Delta m_v$& Sep(${\prime\prime}$) \\
	      \hline
	      HR4414&02 55 39 &11 26 45.3&6.5&1.1&28.5\\
	      HR3174&-09 17 25&08 06 27.4&6.23&1.7&30.8\\
	      HR3428&19 37 50 &08 40 20.7&6.44&1.3&63.2\\
	      HR4752&25 49 26 &12 28 54.7&5.29&1.4&145.4\\
	      HR4884&17 00 33 &12 52 12.2&6.32&0.6&196.5\\
	      HR5010&19 48 03 &13 16 32.1&6.49&1.9&203\\
	      HR4085&02 18 10 &10 24 13.0&6.32&0.3&212\\
	      HR4128&-15 15 43&10 47 37.9&6.67&1.2&74.7\\
	      HR4193&04 44 52 &10 43 20.9&5.79&1.1&6.7\\
	      HR4259&24 44 59 &10 55 36.7&4.5&1.9&6.5\\
	      HR4677&-03 57 14&12 18 09.2&6.99&0.4&20.1\\
	      HR7593&-08 13 38&19 54 37.7&5.71&0.8&35.7\\
	      HR7672&17 04 12 &19 51 17.7&5.8&0.9&203.7\\
	      HR7705&20 53 48 &20 09 56.6&6.48&0.6&83.9\\
	      HR7830&-18 35 00&20 29 53.9&5.94&0.8&21.9\\
	      HR7840&11 15 39 &20 31 13.1&7.11&0.3&16.7\\
	      HR8265&06 37 06 &21 37 43.6&6.18&1.5&39.5\\
	      HR8619&-28 19 32&22 39 44.2&6.31&1.3&86.6\\
	      HR9002&-18 40 41&23 46 00.9&5.29&1&6.6\\
	      HR9044&-27 02 32&23 54 21.4&6.35&0.7&6.4\\
	      HR310 &21 28 24 &01 05 41.0&5.34&0.3&29.9\\
	      HR313 &01 92 24 &01 05 49.1&6.35&1&33\\
	      HR545 &19 17 45 &01 53 31.8&4.83&0.1&7.8\\
	      HR765 &24 38 51 &02 37 00.5&6.5&0.6&38.3\\
	      HR1065&27 34 19 &03 31 20.8&5.96&0.4&11.4\\
	      HR1212&-01 12 15&03 54 17.5&4.79&1.5&6.8\\
	      HR1322&02 51 62 &04 15 28.8&6.31&0.6&64.7\\
	      HR1460&-09 44 12&04 35 14.1&6.37&1&12.8\\
	      HR1505&-08 47 38&04 43 34.7&6.82&0.1&9.3\\
	      HR1600&14 32 34 &04 58 59.4&6.09&1.5&39.3\\
	      HR1610&03 36 58 &05 00 33.9&6.66&0.4&21.3\\
	      HR1619&01 36 32 &05 02 00.0&6.24&1.3&14.2\\
	      HR1753&-18 31 12&05 19 17.4&6.36&0.2&39.3\\
	      HR2174&02 29 58 &06 08 57.8&5.73&1.2&29.3\\
	      HR2356&-07 01 59&06 28 48.9&4.6&0.5&7.1\\
	      HR2948&-26 48 07&07 38 49.3&4.5&0.2&9.9\\
	      HR3010&-14 41 27&07 45 29 &6.07&0.8&16.8\\
	      HR3028&-16 00 52&07 47 45.2&6.43&0.1&130.5\\
	      \hline
	  \end{tabular}
\end{table}

\begin{figure}
	\centering
	\includegraphics[height=8.5 cm]{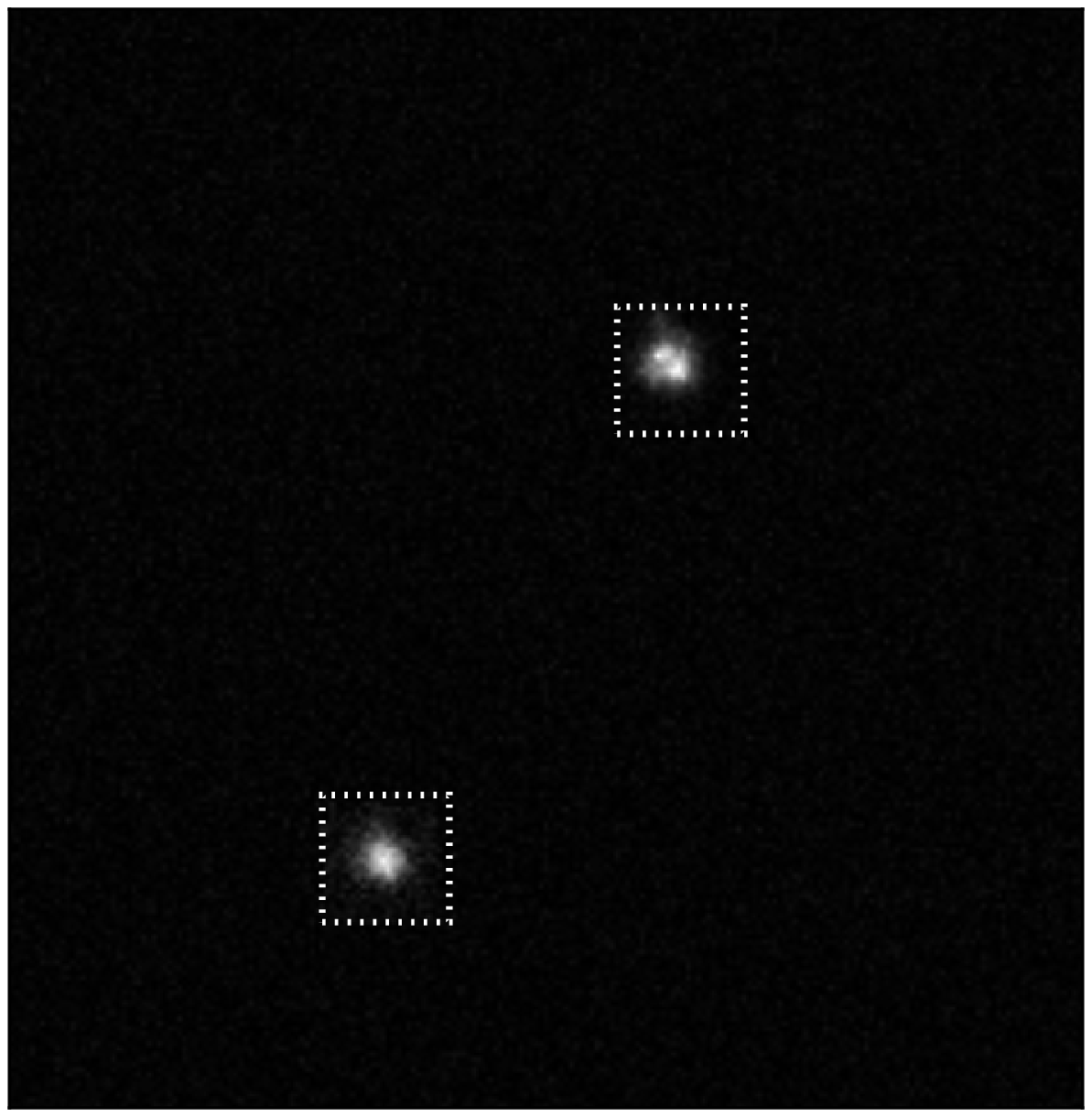}
    \caption{ \label{fig:fig2} 
    A sample frame with 2 target stars observed on 6 April 2015. The object on top-right side (HR4414) has apparent magnitude $m_v$ =~6.5 and bottom-left side (83 Leo B) has $m_v$ =~7.6. The angular separation between two stars is $28.5^{\prime\prime}$. The dotted squares indicate the ROI window around each target.}
\end{figure}

The observations made on selected targets were carried from evening to midnight. The reason for restriction post  midnight is the unprecedented rise in relative humidity (RH) greater than 90$\%$. High RH could cause irretrievable damage to the optics and electronic systems of the telescope. Thus, the observations were confined to below $85\%$ of RH. 

The short exposure images of individual stars within CCD FoV were acquired using ROI technique as illustrated in Figure~\ref{fig:fig2}. The ROI size is varied from $30 \times 30$ pixels to $45\times 45$ pixels. This varied size is chosen to confine the image motion within the specified ROI window. This of course depends upon the prevailing seeing conditions and duration of each observation run.  The exposure time is 8-25 ms and each target pair is observed for 10000-100000 frames over $\approx 4-25$ minutes. 

All our observations are confined to R-band of peak at $ \approx 600$ nm. The observation details are summarized in Table~\ref{tab:observation}. 

\subsection{Data analysis}
  The raw data from CCD is in $\it{.spe} $ format. It is a binary data format with distinct sections of header, data and footer. The header contains offset locations of data, time and duration of observation, ROI position on CCD frame, ROI window size, target of observation etc. Each observation of a target is placed in a single binary data file. Prior to data processing, all the image frames needs to be unstacked.
  
  A code is developed in Python software to unstack each of the image frames. As initial step it could read the number of frames, the offset position, dimensions and number of ROIs. The unstacked frames are available for data processing.
  
  Each of the data frame is cleaned prior to analysis. For this, early sky flats have been captured once in a month to know the QE of the CCD pixels known as flat frames. From the available flat frames we used median count to construct the master flat frame. Dark frames were taken for each observation of the target. A Master dark frame was created by finding the median of dark frames. To minimize the effect of  noise, master dark was subtracted from each data frame and divided by master flat frame. Thus the resultant data was cleaned from bias and flat field errors. 
  
\begin{table}
	\centering
	\caption{Observation details}
	\label{tab:observation}
	\begin{tabular}{lccr} 
	      \hline
		Parameter & Value \\
		\hline
		Wavelength & near R-band (Peak at 600 nm) \\
		ROI size & $30\times30$ to $45\times45$ pixels \\
		Exposure time & 8-25 ms \\
		Magnitude & $m_v < 8$ \\
		Hour angle & $\pm 1$ hour\\
		Declination & $\pm 30^{\circ} $\\
		Air mass & $\leq$ 1.34  \\
		Duration per target & $\approx 4-25$ min.\\
		Frames per target & 10000-100000\\
		Number of nights & 29\\
		Total number of observations & 248\\
		\hline
	\end{tabular}
\end{table}

\subsection{Centroid estimation}

Image motion is estimated by using its centroid. The centroid region is limited to a window of 10 x 10 pixels surrounding the maximum intensity pixel and intensity thresholding is applied. This exercise is aimed to minimize the effect of noise. First order intensity weighted method is used to estimate the centroid of the clean data. 

    \begin{equation}
	X_c = \frac{\sum x_i I_i }{ \sum I_i}, Y_c = \frac{\sum y_i I_i }{ \sum I_i},
	\label {eq:eq1}\\
    \end{equation}    
    \begin{equation}
	\sigma_x = \sqrt[]{\frac{\sum ({X_c-\bar {X_c}})^2 }{n}}, \sigma_y = \sqrt[]{\frac{\sum ({Y_c-\bar {Y_c}})^2 }{n}},
	\label{eq:eq2}\\
    \end{equation}
where $X_c$, $Y_c$ the estimated centroid coordinates of an image, $I_{i,j}$ are pixel intensities, $x_{i,j}$, $y_{i,j}$ are coordinates of pixels, $\bar {X_c}$, $\bar {Y_c}$  mean value of centroids and $\sigma_x$, $\sigma_y$ are the rms error in centroids.

\begin{figure}
    \centering
	\includegraphics[height=8.5cm]{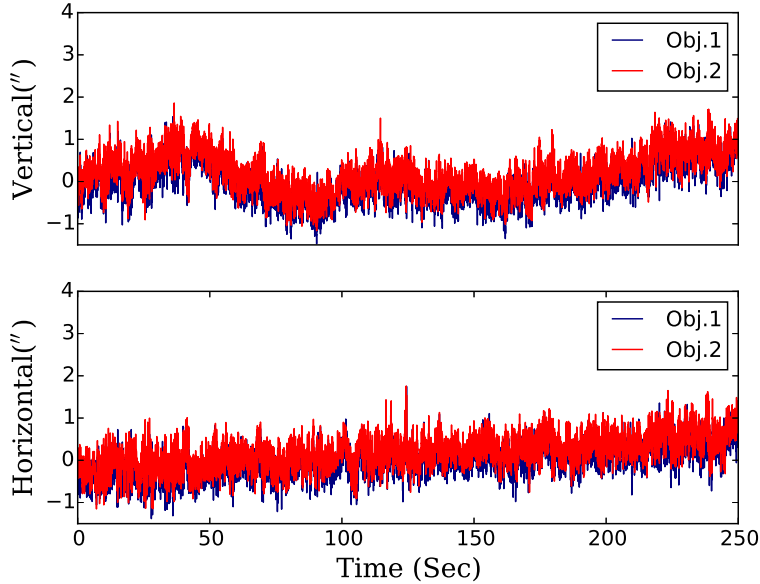}
    \caption{ \label{fig:fig3} 
    Absolute motion of image centroid along horizontal axis (H) and vertical axis (V) of the CCD for object 1 (HR4414) and object 2 of the target pair. In this image X-axis is duration of the observation in seconds and Y-axis is centroid motion of the objects in arc seconds ($^{\prime\prime}$).}
\end{figure}

In Figure~\ref{fig:fig3},  the centroid motion of star 1 (HR4414) and star 2 along horizontal axis (H) and vertical axis (V) of the CCD is plotted. Both objects were observed simultaneously. 

\section{Estimation of seeing}

The Fried's parameter $r_0$ is a single parameter used to represent the turbulence strength (\citealt{Fried66}). It is defined as the spatial scale at which the {\it rms} phase variation is  one radian in the distorted wave-front. It is dependent on the refractive index structure constant of turbulent medium, wavelength  and zenith angle of the observations. Refractive index structure constant is integrated over multiple layers of the atmosphere and it varies randomly with time. This causes the continuous random fluctuations in $r_0$. Instantaneous variations of this parameter causes the random motion and blurring of the image. In practice, the image quality is expressed as $\lambda/r_0$, also known as the {\it seeing}.

\subsection{From {\it rms} image motion}

\begin{figure}
    \centering
	\includegraphics[height=8.5cm]{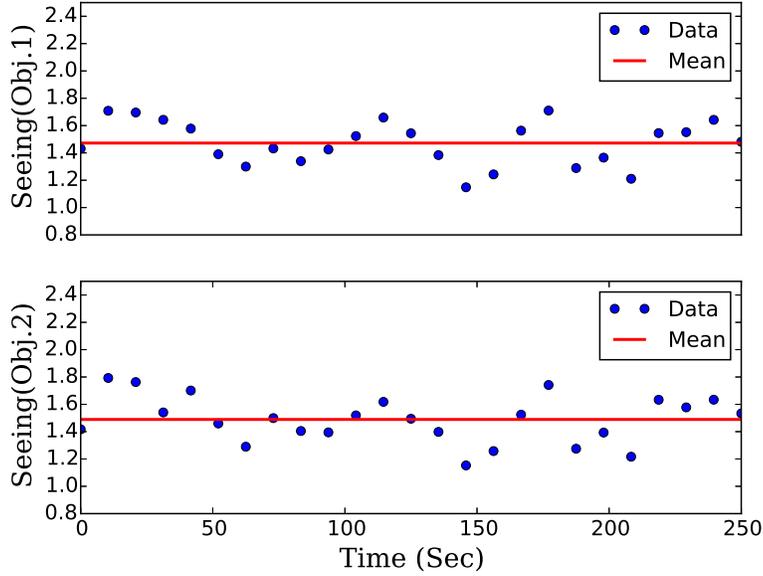}
    \caption{ \label{fig:fig4} 
    Atmospheric seeing measured from the root mean square ({\it rms}) data of the centroids of HR4414 object 1 (top) and object 2 (bottom). Seeing ($^{\prime\prime}$) is measured over every 10 seconds. The bold line is the mean seeing. The measured mean seeing is $\approx 1.49^{\prime\prime} \pm 11\%$ for both objects.}
\end{figure}

We estimated the {\it seeing} from the motion exhibited by a burst of short exposure images. Essentially, $r_0$ was estimated from the images using Equation~\ref{eq:eq3} (\citealt{mar87}).

   \begin{equation}
	\frac{r_{0}}{1m}= \bigg[\frac{0.0431}{\sigma} \frac{\lambda}{0.5\mu} \bigg(\frac{d}{1m}\bigg)^{-\frac{1}{6}}\bigg]^\frac{6}{5}.
	\label{eq:eq3}\\
   \end{equation}   
Here, $\sigma$ is root mean square of the image motion, estimated over 10~s interval of data.
As a result,this expression gives $r_0$ for every 10 s and mean $r_0$  is calculated for the entire duration of  each observation. The estimated $r_0$ (cm) is converted to atmospheric seeing ($^{\prime\prime}$). 
In Figure~\ref{fig:fig4}, the estimated seeing for object HR4414 is plotted. The mean seeing measured from this observation is $1.49^{\prime\prime} \pm 11\%$.

\subsection{From FWHM}
The mean seeing measured from the aforesaid method is compared with the full width half maximum (FWHM) of the long exposure image. In this case the long exposure image is obtained by co-adding the short exposure images as they have negligible time interval between them.  The FWHM of a long exposure image is a standard estimation of atmospheric seeing conditions.  Assuming negligible tracking errors within a time span of 10s, one can estimate FWHM every 10 seconds and hence the mean seeing for full length of the observation.  The FWHM is related to $r_0$ as $0.98{\lambda}/{r_0}$ and $r_0$ (cm) is converted to seeing ($^{\prime\prime}$). The measured seeing for the object HR4414 is $1.54^{\prime\prime} \pm 9\%$.
 
\subsection{From the Spectral-ratio}
Estimation of $r_0$ using the spectral ratio method was formulated by \cite{VonDer84}.
In this method the squared modulus of ensemble averaged Fourier transform of an image (${S_i}(\bar {q})$) is divided by the ensemble averaged power spectrum of that image. The observed radial profile is compared with theoretical model to estimate $r_0$. 
The expression for the spectral-ratio method reads as follows. 
 \begin{equation}
  \varepsilon({\bar q}) = \frac{{\vert\langle {S_i}(\bar {q})\rangle\vert}^2}{\langle{\vert {S_i}(\bar {q})\vert}^2 \rangle}
  \label{eq:spectra}\\
 \end{equation}
 \begin{equation}
  E[\varepsilon({\bar q})] = \frac{E[{\vert\langle {S_i}(\bar {q})\rangle_{SE}\vert}^2]}{E[\langle{\vert {S_i}(\bar {q})\vert}^2 \rangle]}
  \label{eq:spectramodel}\\
 \end{equation}
 In Equation~\ref{eq:spectramodel}, E( ...) denotes an ensemble average.

 The radial profiles obtained from observational data and theoretical model have been over plotted to estimate seeing. The model is simplified and it is expressed in terms of q and $\alpha$, where q is wave number and $\alpha$ is modified Fried's parameter. The wave number is defined as the ratio of spatial frequency and theoretical cut off frequency of the telescope $f_c= \frac{D}{\lambda R}$, whereas D is diameter, R is focal length of the telescope and $\lambda$ is light wavelength. The modified Fried's parameter $\alpha$ is given as ${r_0}/D$.  
  \begin{equation}
  \alpha=Aq^{B} \ \forall \ {\alpha \le 0.3}
  \label{eq:alplaqrel}
 \end{equation}
 The coefficients A and B have been chosen such that the falling end of the radial profiles of the observational and theoretical models will coincide. The $r_0$ of observational data is obtained by finding the better represented theoretical radial profiles (~\citealt{VonDer84}). 

\begin{figure}
    \centering
	\includegraphics[height=7.5 cm]{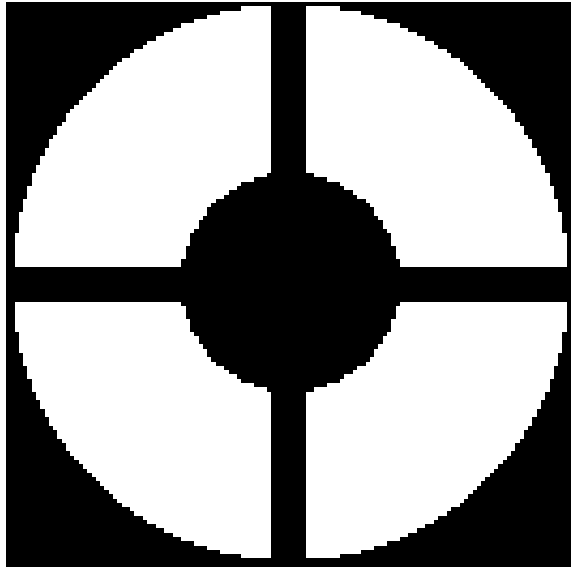}
 \caption{	\label{fig:fig5}
	 The aperture geometry used in simulating the theoretical transfer functions and estimating  the spectral ratio constants. The black regions indicate central obscuration and spiders holding the secondary. The white region indicates the useful aperture area.}
\end{figure}

 We estimated $r_0$ using this method after incorporating the following two changes: First, we estimated the library of short exposure transfer functions and the speckle transfer functions numerically for our annular aperture geometry (c.f. Figure~\ref{fig:fig5}) assuming Kolmogorov turbulence model atmosphere. We then  estimated the spectral ratio constants  A \& B using the numerically simulated spectral ratio.
 
 After estimating the theoretical spectral ratio constants for our annular aperture, we used them to estimate $r_0$ for our observed data. For the object HR4414 the mean was found to be $1.71^{\prime\prime} \pm13\%$. 
 
\begin{figure}
    \centering
	\includegraphics[height=6.5cm]{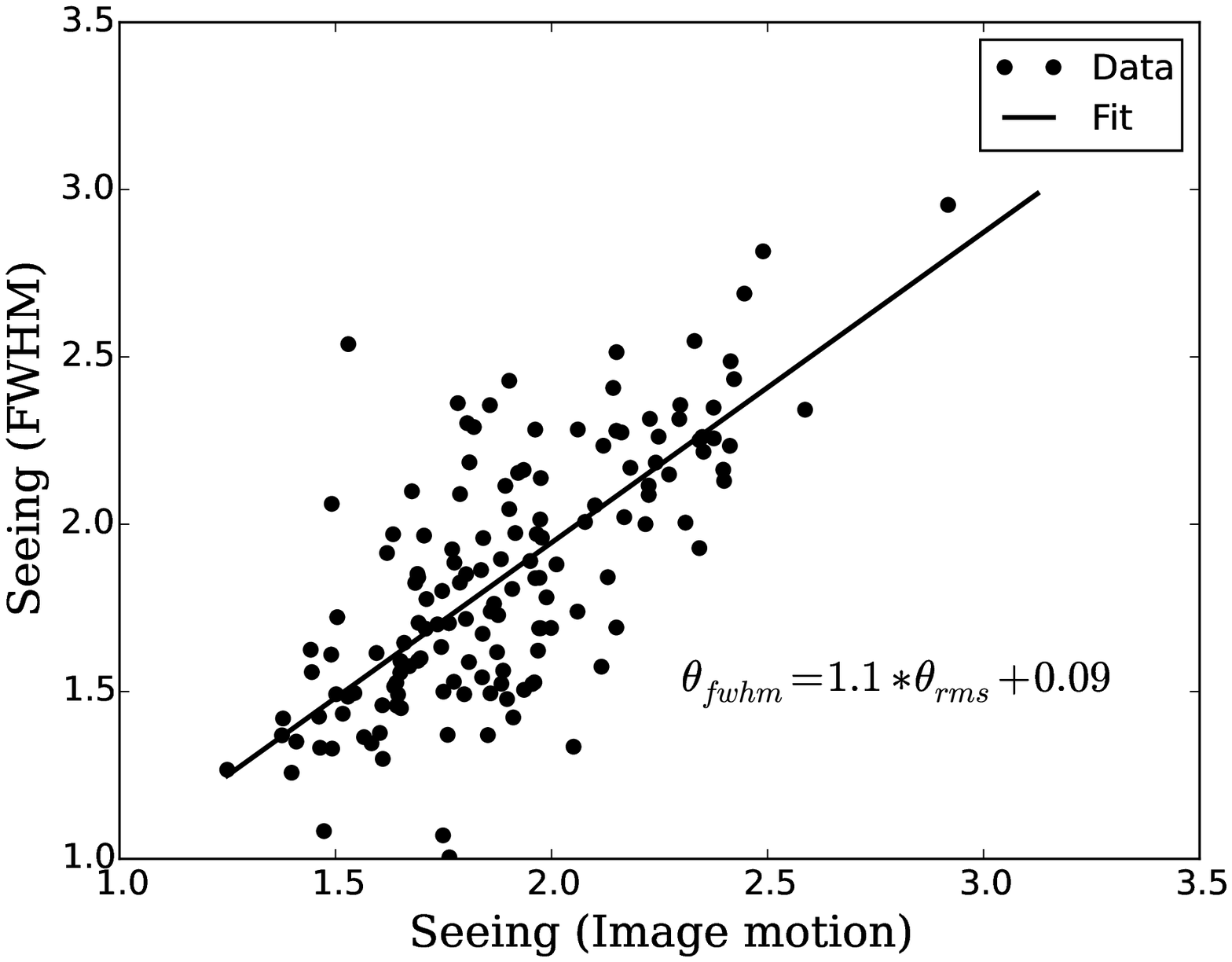}\\
	\includegraphics[height=6.5cm]{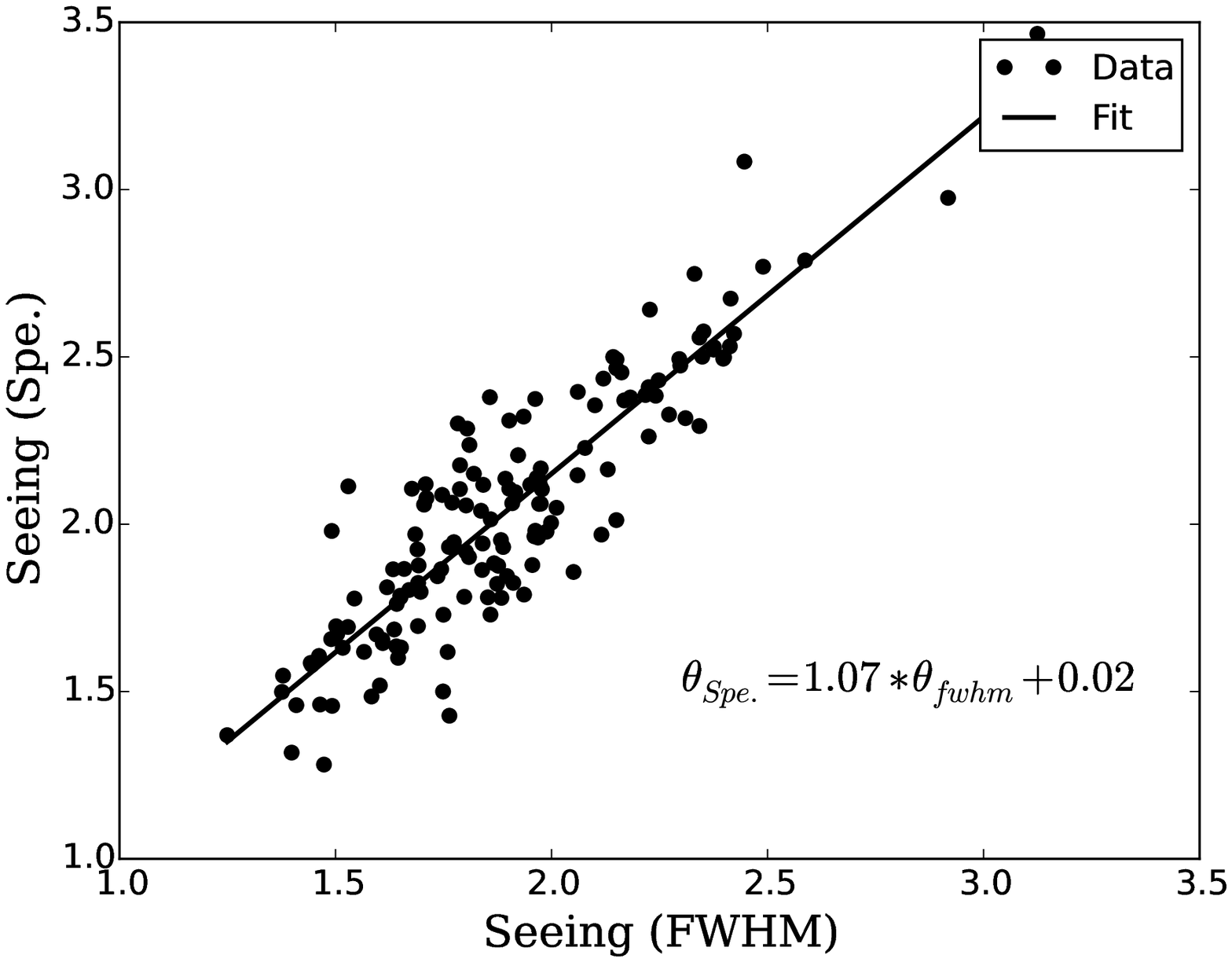}\\
	\includegraphics[height=6.5cm]{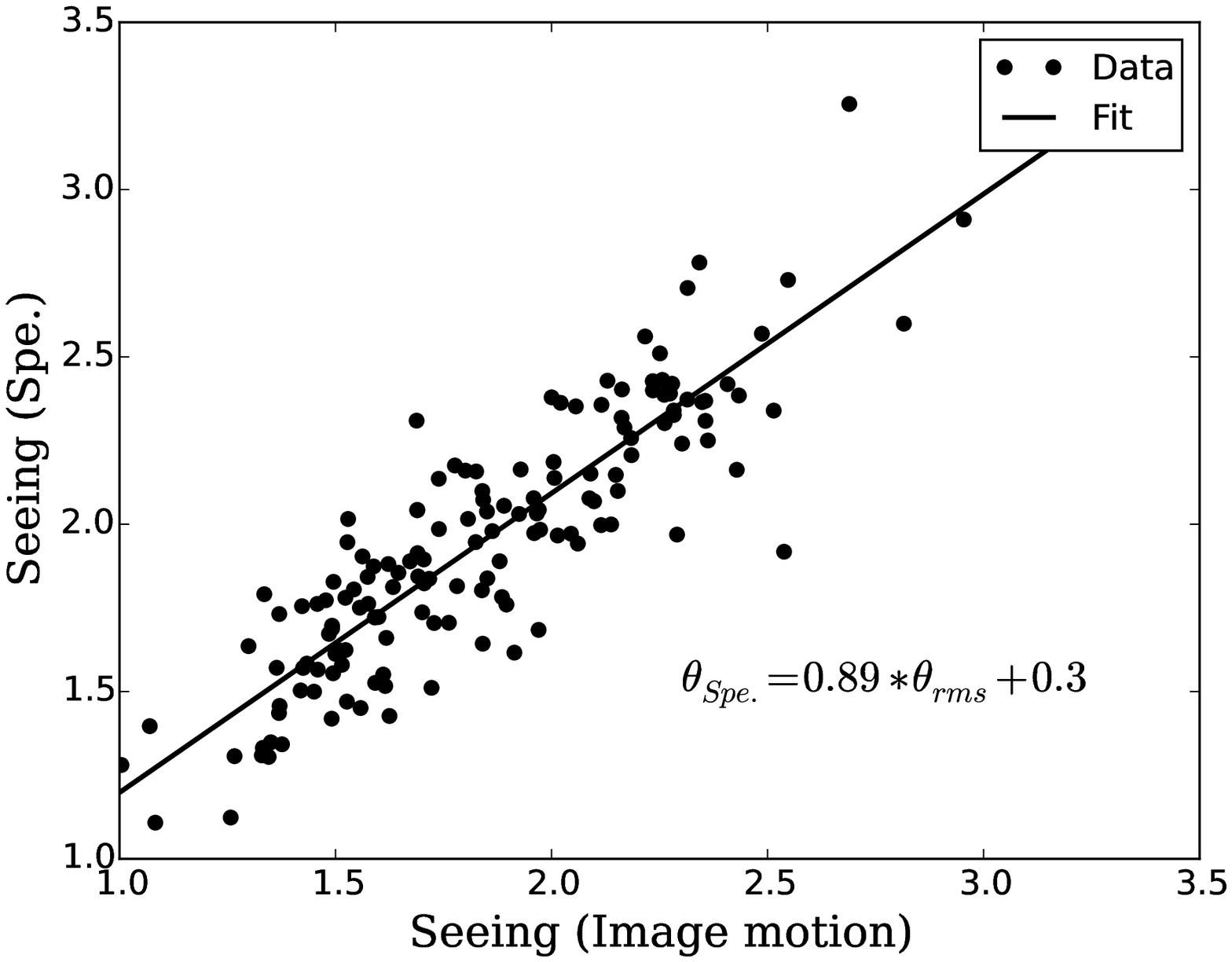}\\
    \caption{\label{fig:fig6} 
    Seeing measured from centroid motion of object ({\it rms}), FWHM and spectral ratio (Spe.) methods has been compared. The units of the axis are in arc seconds. The data points in this plot are seeing measured for the targeted objects. The solid line is the approximated relation between seeing measured from both methods. A linear model has been fit into the data obtained and relation between them is shown in the plot.}
\end{figure} 

\subsection{Comparison of \texorpdfstring{$r_0$}{Lg}}
The seeing estimated using the image motion method, the FHWM and the spectral ratio methods have been compared and the results obtained from these methods have been presented in Figure~\ref{fig:fig6}.  It is observed that the seeing estimated 
from the aforesaid  methods has correlation of $\approx 86\%$(mean of the three correlations). Our observation is that the $r_0$ estimated from the angle of arrival fluctuation method is higher than that from the spectral ratio method and it is consistent with what has been reported in the literature elsewhere (\citealt{Goode2000}). However, the exact ratio in our case is 1.14 and it is less than that reported by  \cite{Goode2000}. 
As there is good correlation between the values estimated from different methods, hereafter we refer to the value estimated from the $rms$ image motion in our discussions.

\subsection{Measurement of \texorpdfstring{$r_0$}{Lg} over long-term}
\begin{figure}
    \centering
	\includegraphics[height=6.5cm]{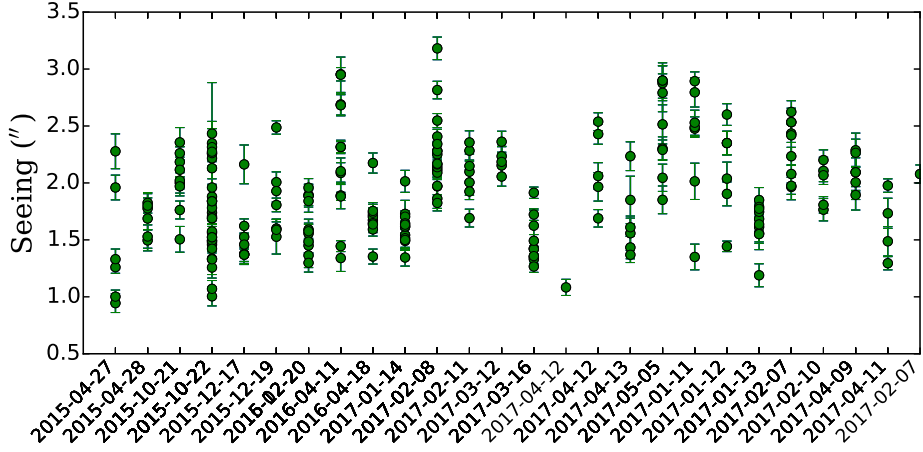}
    \caption{\label{fig:fig7} Each of the data point with error is mean seeing ($^{\prime\prime}$) estimated for each observation. The plot is the result of 248 observations of more than 50 targets observed in 29 nights over a period of 2-years. On average the error in each estimation is $\approx \pm 11\%$. The vertical projection of data points is the mean seeing of the observations during same night. The projection is as high as $1.5^{\prime\prime}$.}
\end{figure}

\begin{figure}
    \centering
	\includegraphics[height=8.5cm]{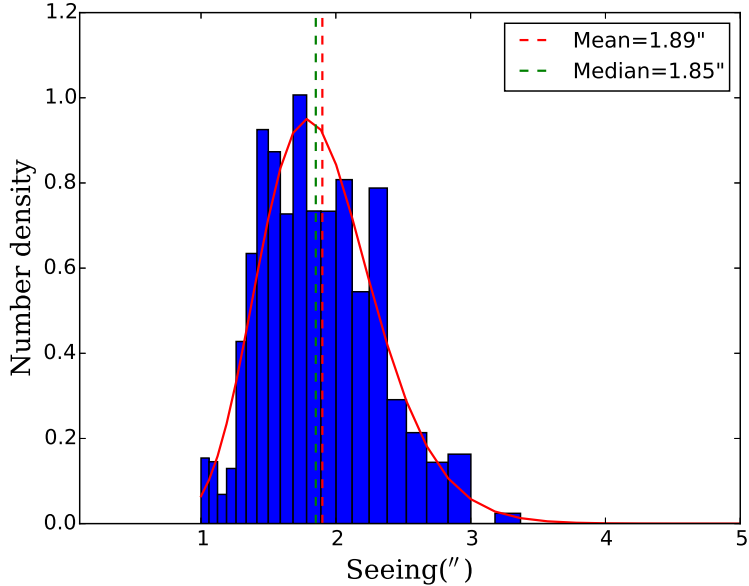}
    \caption{	
    \label{fig:Histogram}
Probability density of seeing values and log-normal distribution. The mean and median seeing observed from the data are $1.89^{\prime\prime} \pm 11\%$ and $1.85^{\prime\prime}$ respectively.}
\end{figure}

In Figure~\ref{fig:fig7}, seeing estimated from {\it rms} image motion is plotted against the civil day of observation. It contains  mean seeing from 248 observations on more than 50 targets ( including the list shown in Table~\ref{tab:Targets}) observed over 2 years. The vertical line along any night  shows the temporal variation of the seeing over the time period of observations. This is because the atmospheric turbulence causes temporal fluctuations in $r_0$. It varies with time, position of the target and depends on several observational conditions. The mean seeing of the telescope site is evaluated as the statistical average of the estimated seeing over the entire duration of observations. It is estimated as $1.89^{\prime\prime} \pm 11\%$. The probability density of observed seeing is plotted in Figure~\ref{fig:Histogram}. The normalized distribution of the data is plotted over the histogram. The median seeing at the telescope site is observed as $\approx 1.85^{\prime\prime}$

\section{Estimation of tilt-anisoplanatic angle}

To estimate $\theta_0$, several pair of stars separated by an angle $\theta$, at the image plane, have been considered. 
Table~\ref{tab:Targets} indicates the targets selected for the estimation. Each of the star pair with ROIs is observed simultaneously with an exposure time of 25 ms.  The limitation on exposure time is due to minimum time required to readout CCD frame (2 ROIs with $40\times40$ pixels). The duration of observation of  each target is $\approx 4$ minutes,  and it corresponds to about 10000 frames. To quantify the relation in their image motion, the centroids of the images are correlated. An expression that is used for the measurement of correlation coefficient ($\rho$) is shown in Equation~\ref{eq:eq5}. This parameter is measured for several pair of stars with different angular separation. 

    \begin{equation}
	\rho_x = \frac{\sigma_{x1x2}}{\sqrt{\sigma_{x1}\sigma_{x2}}}, \rho_y=\frac{\sigma_{y1y2}}{\sqrt{\sigma_{y1}\sigma_{y2}}}.
	\label{eq:eq5}\\
    \end{equation}    
In the above equation $\sigma_{x1x2}$, $\sigma_{y1y2}$ is the covariance of the image centroid motions of object 1 and object 2 along H and V axis respectively, $\sigma_{x1}$,$\sigma_{x2}$, $\sigma_{y1}$, $\sigma_{y2}$ are image motion variance (same as Equation~\ref{eq:eq2}) of object 1 \& 2 and $\rho_x$, $\rho_y$  are correlation coefficients of the image centroid motion along the two orthogonal axis. The estimated $\rho$ is the mean of $\rho_x$ and $\rho_y$. 

\begin{figure}
    \centering
	\includegraphics[height=8.5cm]{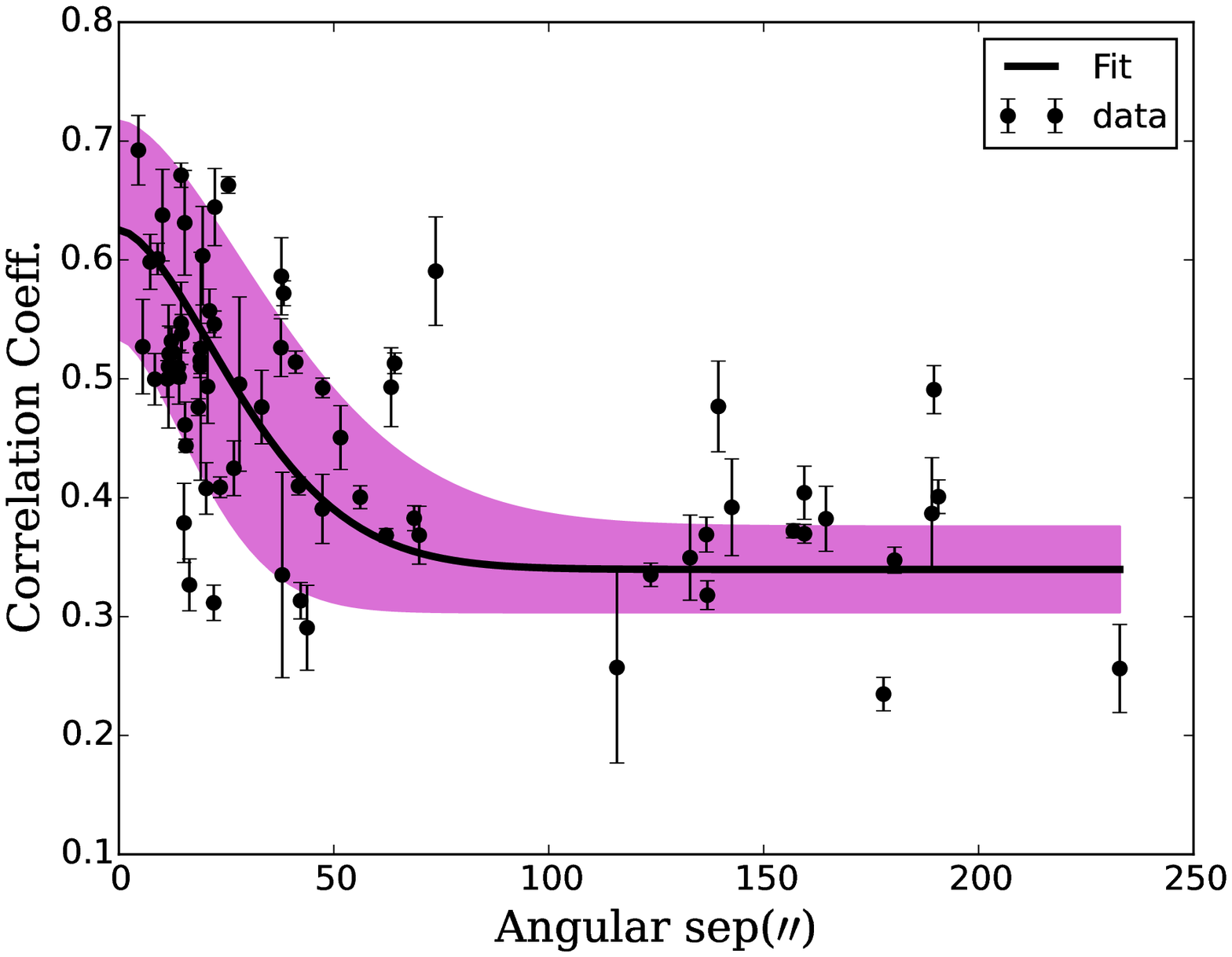}
    \caption{    
    \label{fig:fig8}
    The relation between the image motion correlation between stars of a target with respect to the angular separation between them. The plot has 73 observational data points.  The X-axis and Y-axis are the angular separation in arc seconds ($^{\prime\prime}$) and correlation coefficient respectively. The data has an estimated error as high as $\approx 21\%$. Approximately, $60 \% $ and above of correlation is present at angular separation of $12^{\prime\prime}$. Thick band in orchid color is $5\sigma$ level of error estimation. The empirical relation between $\rho$ and $\theta$ is mentioned.}
\end{figure} 

The observational data is shown in the Figure~\ref{fig:fig8} where $\rho$ is plotted against the $\theta$. It is evident that the $\rho$ is inversely related to $\theta$. Even though image motion correlation is  largely dependent on $\theta$, it is also influenced by seeing and wind speed during the observation and zenith angle of the target. The effects of these parameters on the measurement of $\rho$ have been present in the image centroid motion. Thus the measured value of $\rho$ is not purely due to $\theta$ rather it is the integrated effect of all these parameters.

\begin{equation}
 \rho=a_1+a_2  \exp[-(\theta/\theta_0)^{5/3}].
 \label{eq:eq6}
 \end{equation} 
 The $\theta_0$ is obtained from an empirical relation derived from the statistical relation between $\rho$ and $\theta$. To obtain such relation, an analytical expression (Equation~\ref{eq:eq6}), based on Kolmogorov turbulence model, has been fit to the observational data. In Equation~\ref{eq:eq6}, $\rho$ is correlation coefficient, $a_1$, $a_2$ coefficients and $\theta$, $\theta_0$ are angular separation and isoplanatic angle respectively. 
 A least square approximation model is fit to the data to estimate these parameters. From this approximation it is estimated that $a_1 \approx 0.34$, $a_2 \approx 0.29$ and $\theta_0 \approx 36^{\prime\prime}$. 
 
 The fitted data plot is drawn in solid line in the Figure~\ref{fig:fig8}. The thick band in orchid color shows the 5 sigma level of error estimation. Most of the data fall within this limit. The figure shows the image motion correlation between two objects has declining trend with increase in angular separation. Approximately, $60\%$ of correlation has been observed down to angular separation of $12^{\prime\prime}$. The correlation coefficient has reached $\approx 44\%$  at  the separation of $\approx 36^{\prime\prime}$. This model equation is valid for objects with angular separation ($\theta$) $\gtrsim$ $6.4^{\prime\prime}$, as is the case with our data set.

\section {Estimation of coherence time}
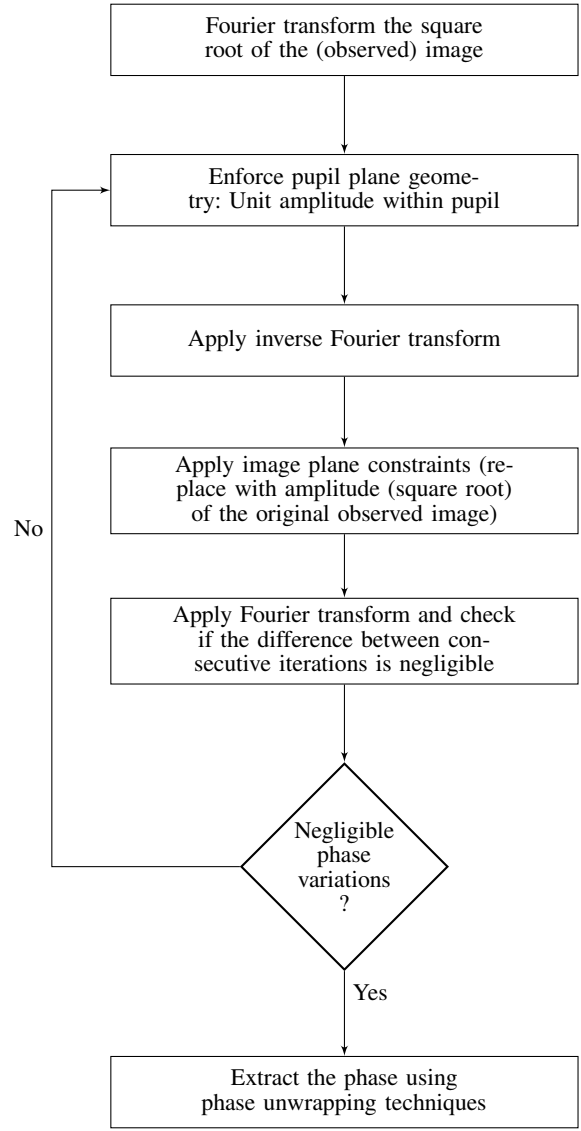
\begin{figure}[!htb]
    \centering
    \begin{tikzpicture}[
      >=latex',
      auto
    ]
      \node [intg] (kp)  {Fourier transform the square root of the (observed) image};
      \node [intg]  (ki1) [node distance=2 cm ,below of = kp] {Enforce pupil plane geometry: Unit amplitude within pupil};
      \node [intg]  (ki2) [node distance=2 cm ,below of = ki1] {Apply inverse Fourier transform};
      \node [intg] (ki3) [node distance=2 cm,below of=ki2] {Apply image plane constraints (replace with amplitude (square root) of the original observed image)};
      \node [intg] (ki4) [node distance=2 cm,below of=ki3] {Apply Fourier transform and check if the difference between consecutive iterations is negligible};
       \node [decision] (ki5) [node distance=3 cm,below of=ki4] {Negligible phase variations ?};
      \node [intg] (ki6) [node distance=3 cm,below of=ki5] {Extract the phase using phase unwrapping techniques};

      \draw[->] (kp) -- (ki1);
      \draw[->] (ki1) -- (ki2);
      \draw[->] (ki2) -- (ki3);
      \draw[->] (ki3) -- (ki4);
      \draw[->] (ki4) -- (ki5);
      \draw[->] (ki5.west) -- ($(ki5.west)+(-2.5,0.0)$) |- node [near start] {No} (ki1);
      \draw[->] (ki5.south) -- node [near start] {Yes} (ki6);
    \end{tikzpicture}
    \caption{Flow chart to extract phase-fronts from observed images. \label{Fig:FlowChart}}
\end{figure}

We adopted the following procedure to estimate the atmospheric coherence time ($\tau_0$) from the series of short exposure images. The general idea is to extract a series of phase-fronts from the series short exposure images.
We make use of the Fourier transform relationship between the image plane the telescope pupil plane and apply the Gerchberg-Saxton algorithm (\citealt{Gerch72}) to recover, iteratively,  the complex phase distribution in the pupil plane. We then extract  the phase from this complex distribution using phase-unwrapping algorithm (\citealt{phaseunwrap}). The iterative procedure is shown in Figure~\ref{Fig:FlowChart}.

The $\it{rms}$ phase variations of the pupil plane phase $\phi$ corresponding to each frame is estimated.
We define temporal phase structure function $D_{\phi}$ as given in Equation~\ref{eq:eq9} and model the same as in Equation~\ref{eq:eq10}.
    \begin{equation}
	{D_\phi}(\tau) = \langle |{\phi}(t)-{\phi}(t+\tau)|^2 \rangle,
        \label{eq:eq9}\\
    \end{equation}
    \begin{equation}
	{D_\phi}(\tau) = \bigg(\frac{\tau}{\tau_0}\bigg)^{\beta},
	\label{eq:eq10}\\
    \end{equation}
where the angular brackets indicate ensemble average, $\tau$ is time to acquire an image, $\tau_0$ is the coherence time and $\beta$ is a constant.  
 
 To simplify the model fitting, Equation~\ref{eq:eq10} is rewritten in logarithmic scale. The resultant linear equation is given below.
 
 \begin{equation}
  log[D_{\phi}(\tau)] = {\beta}[log(\tau)-log(\tau_0)].
  \label{eq:eq11}
 \end{equation}
Both $\beta$ and $\tau_0$ are estimated from the model.

\begin{figure}
    \centering
	\includegraphics[height=8.5cm]{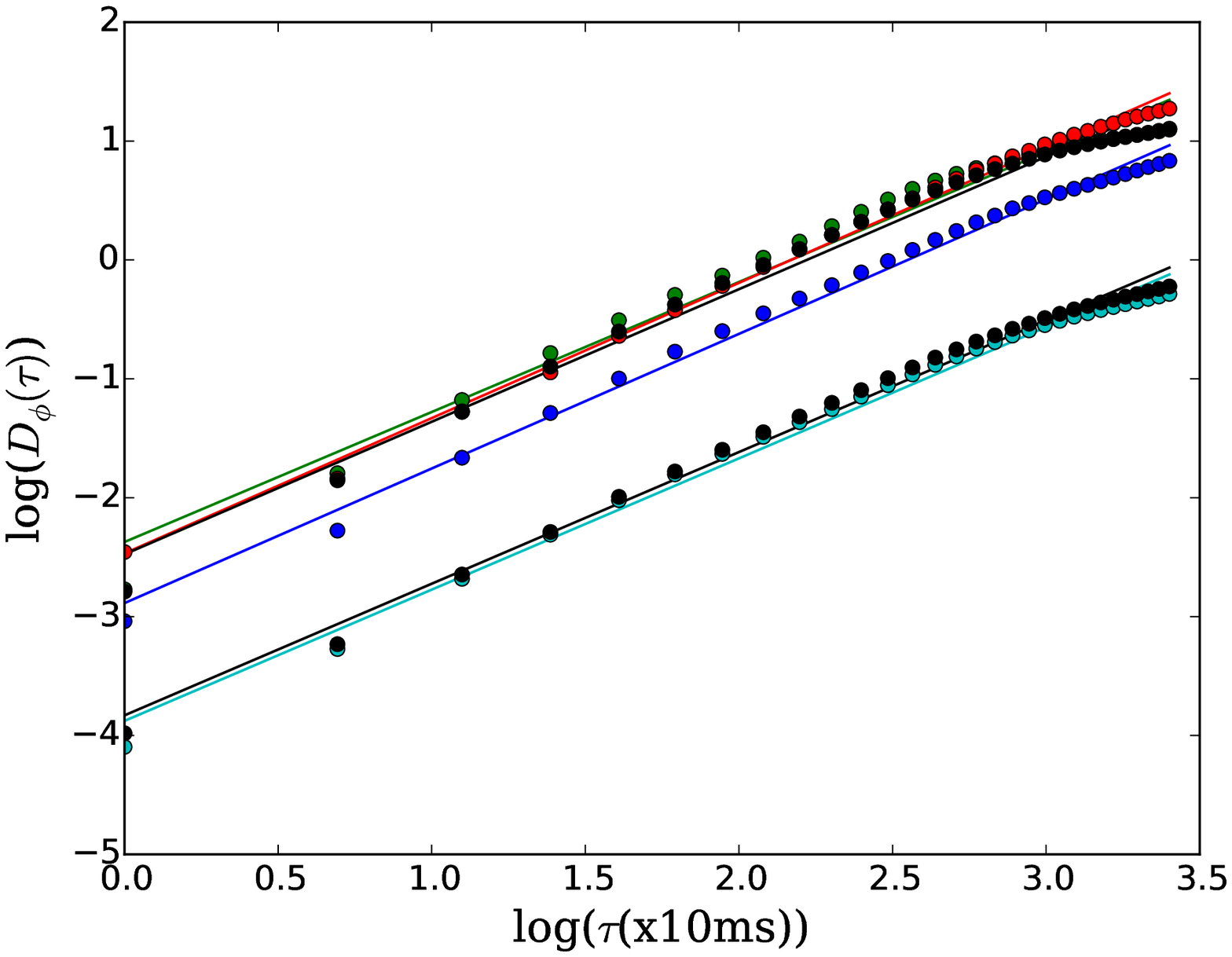}
    \caption{    
    \label{fig:fig9}
    Log amplitude of temporal phase structure function. It is ensemble average of the rms phase variations of star images with time. The X-axis is time and the Y- axis is log amplitude of the structure function. The data points in the plot are obtained the observational data and the solid lines are model fit to the structure function.}
\end{figure}

The theoretical value of $\beta$ is ${5}/{3}$ for Kolmogorov Turbulence.  
 $\beta$ \& $\tau_0$   estimated for a few representative cases, by fitting the model to the estimated structure functions (Figure~\ref{fig:fig9}). 
\begin{table}
	\centering
	\caption{Turbulence coherence time estimation. It is measured for different target during the same night. The presented data is of the observation conducted on 27 April 2015. The data contains the power value of the structure function and the coherence time. The mean coherence time is $\sim$2.4 ms}
	\label{tab:Ctime}
	\begin{tabular}{lccr} 
	\hline
	Object & $\beta$ & $\tau_0$ (ms) \\
	\hline
	  HR4128 &1.13& 1.49  \\
	  HR4414 &1.10& 3.54 \\
	  HR4752 &1.14& 2.53  \\
	  HR4884 &1.12& 2.47 \\
	  HR4085 &1.10& 2.42 \\
	  HR5010 &1.02& 1.90 \\
	\hline
	\end{tabular}
\end{table}
are  provided in Table~\ref{tab:Ctime}. It indicates  that the coherence time is varying between 1.5 ms and 3.5 ms and $\beta$ is varying between 1 and 1.4.  

\section{Discussions}
We have estimated a median atmospheric seeing of $\approx1.85^{\prime\prime}$, tilt-anisoplanatic angle of $\approx36^{\prime\prime}$ for $44\%$ image motion correlation and a mean coherence time of $\approx2.4$ ms. In what follows, we shall discuss how these values play a decisive role in designing AO system for our 1.3 m telescope at VBO.
Particularly, we will discuss how the knowledge of these parameters helps to specify the stroke needed for the deformable mirror, the temporal frequency required for the data acquisition  and the effectiveness of the AO correction from the lock position.

We shall arrive at the stroke required, assuming that the AO system should be operable when $r_0$ is as low as 5.5 cm. The residual {\it rms} wave-front error at 600 nm in the instantaneous wave-front (i.e. excluding the tip \& tilt) is shown in Equation~\ref{eq:eq12} (\citealt{tyson98}).

\begin{equation}
\sigma = 0.366\frac{\lambda}{2\pi}\bigg(\frac{D}{r_0}\bigg)^{5/6} = 0.512 \mu m.
\label{eq:eq12}
\end{equation}

 Assuming a 5-$\sigma$ level for the instantaneous wave-front, the peak wave-front error is
 2.56 $\mu$m. It should be noted that with the reflective geometry, the actual stroke needed will be half of this value. Thus, a deformable mirror with maximum stroke of $\sim$ 3 $\mu$m should be conservative upper limit.
 If we specify 0.1 Strehl ratio at 600 nm, the residual mean square wave-front phase error is 2.3 radian$^2$. Assigning 1/3 weight to the wave-front fitting error, we arrive at the number of actuators N using the expression given by \cite{har98}.
 
 \begin{equation}
\sigma_{fitting} = 0.3N^{-5/6}(D/r_0)^{5/3}.
\label{eq:eq13}
\end{equation}
Substituting the values, we get N$\sim$183. Thus, we would require about 180-200 actuators in our system.

The wave-front correction needs to be applied within one coherence time $\tau_0$. The closed loop correction bandwidth is $\sim$ $\frac{1}{2\pi\tau_b}$  where $\tau_b$ is the time gap between the wave-front sensing and the wave-front correction. Assuming a factor of six to  ten times the closed loop correction bandwidth is required for the AO loop, we arrive at the loop frequency of $\sim$ 400-600 frames per second. 

Tilt-anisoplanatic angle $(\theta_0) $ is an important parameter for natural guide star AO systems. As discussed in the introduction, the better choice of reference object is its proximity to target, so that the AO system performance will be effective. Thus it limits the effectively corrected field of view for observations. In our case, when we use a bright reference object nearby our target of interest, their angular separation should be less that 36$^{\prime\prime}$ so that at least the tilt component of the wave-front correction will be effective.
\section{Summary and Conclusion}
\begin{enumerate}
    \item  We have estimated $r_0$ (alternatively the seeing) from 248 distinct observations spanning over 29 days. The made use of the fast CCD images and measured $r_0$ from angle of arrival fluctuations. We compared the estimated values with those estimated from other methods  and found that the values match with 14\% uncertainty.
    \item With the limited data, the median seeing at 600 nm is found to be 1.85$^{\prime\prime}$, the tilt-anisoplanatic angle $\theta_0$ is 36$^{\prime\prime}$ for 44\% correlation, and the atmospheric coherence time, estimated for six different observations, is $\sim$ 2.4 ms.
    \item The estimated parameters should be considered as preliminary, as the values are likely to change with further data. Nevertheless, it helps in identifying the design parameters of the adaptive optics system to be built on this telescope.
    \item The observed values reinforce the need to build an adaptive optics system for achieving diffraction limited resolution. An infra-red AO system will be preferable as the coherence time is likely to be better than what has been observed in the near R-band.
\end{enumerate}

\section*{Acknowledgements}
We would like to thank Mr. P. Anbhazagan, field officer in-charge of Kavalur observatory for his valuable support and observing assistants V. Moorthy, G. Selvakumar and S. Venkatesh for their help during the course of this work.

\bibliographystyle{raa}
\bibliography{ms2018-0256article}
\end{document}